\documentclass{sigchi-ext}
\usepackage[T1]{fontenc}
\usepackage{textcomp}
\usepackage[scaled=.92]{helvet} 
\usepackage{graphicx} 
\usepackage{balance}  
\usepackage{booktabs} 
\usepackage{ccicons}  
\usepackage{ragged2e} 
\usepackage[ampersand]{easylist}

\def\plaintitle{SIGCHI Extended Abstracts Sample File: Note Initial
  Caps} 
\def\emptyauthor{}
\def\plainkeywords{Reminiscence; Social interactions; Older adults; Residential care; Design recommendations}

\title{Designing for Co-located and Virtual Social Interactions in Residential Care}

\numberofauthors{4}
\author{%
  \alignauthor{%
    \textbf{Francisco Ibarra}\\
    \affaddr{University of Trento} \\
    \affaddr{Trento, Italy} \\
    \email{fj.ibarracaceres@unitn.it} }
  \alignauthor{%
    \textbf{Francesca Fiore}\\
    \affaddr{University of Trento}\\
    \affaddr{Trento, Italy}\\
    \email{francesca.fiore@unitn.it} } \vfil
  \alignauthor{%
    \textbf{Marcos Baez}\\
    \affaddr{University of Trento} \\
    \affaddr{Trento, Italy} \\
    \affaddr{Tomsk Politechnic University}\\
    \email{baez@disi.unitn.it} }\alignauthor{%
    \textbf{Fabio Casati}\\
    \affaddr{University of Trento}\\
    \affaddr{Trento, Italy}\\
    \affaddr{Tomsk Politechnic University}\\
    \email{fabio.casati@unitn.it} } \alignauthor{%
} }

\definecolor{linkColor}{RGB}{6,125,233}
\hypersetup{%
  pdftitle={\plaintitle},
  pdfauthor={\emptyauthor},
  pdfkeywords={\plainkeywords},
  bookmarksnumbered,
  pdfstartview={FitH},
  colorlinks,
  citecolor=black,
  filecolor=black,
  linkcolor=black,
  urlcolor=linkColor,
  breaklinks=true,
}

\begin{document}

\CopyrightYear{2018} 
\setcopyright{rightsretained} 
\conferenceinfo{DIS'18 Companion}{June 9--13, 2018, , Hong Kong}\isbn{978-1-4503-5631-2/18/06}
\doi{https://doi.org/10.1145/3197391.3205424}

\copyrightinfo{\acmcopyright}

\maketitle

\RaggedRight{} 

\begin{abstract}

In this paper we explore the feasibility and design challenges in supporting co-located and virtual social interactions in residential care by building on the practice of reminiscence. 
Motivated by the challenges of social interaction in this context, we first explore the feasibility of a reminiscence-based social interaction tool designed to stimulate conversation in residential care with different stakeholders. Then, we explore the design challenges in supporting an assisting role in co-located reminiscence sessions, by running pilot studies with a technology probe. 
Our findings point to the feasibility of the tool and the willingness of stakeholders to contribute in the process, although with some skepticism about virtual interactions. The reminiscence sessions showed that compromises are needed when designing for both story collection and conversation stimulation, evidencing specific design areas where further exploration is needed.

\end{abstract}


\category{H.5.m.}{Information Interfaces and Presentation
  (e.g. HCI)}{Miscellaneous} {}{}

\section{Introduction}

Failing to maintain meaningful connections and to stay socially active is a growing concern in residential care \cite{theurer2015need}. 
Transitioning to residential care requires adjustments difficult to assimilate, by both older adults and their families \cite{lee2002review}.
In turn, this requires nursing homes (NH) to devise special activities that promote social integration and a sense of community. 
Despite these efforts, social isolation and loneliness still plague residential care \cite{bekhet2012mental,nyqvist2013social}, with devastating effects on the quality of life of older older adults \cite{brummett2001characteristics}.

\begin{marginfigure}[0pc]
  \begin{minipage}{\marginparwidth}
    \centering
  \includegraphics[width=.93\columnwidth]{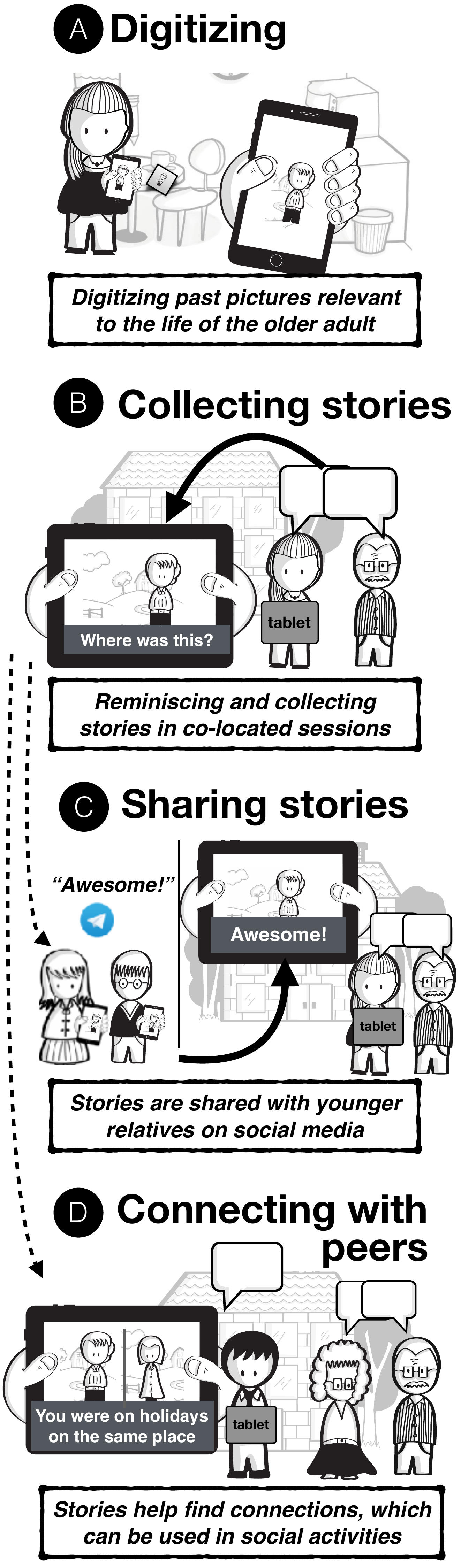}
  \captionof{figure}{Concept explained}
  \label{figure:collage}
  \end{minipage}  
\end{marginfigure}

In this paper we explore the feasibility of reminiscence technology to support and stimulate co-located and virtual interactions in residential care. While the idea of reminiscence technology is not novel (see \cite{lazar2014systematic} for a review), and previous work have taught us valuable lessons on how to facilitate usage by older adults \cite{piper2013audio,west2007memento}, stimulate cognitive functions \cite{lee2014picgo,subramaniam2016digital}, collect memories and support storytelling \cite{astell2009working,lee2014picgo,lee2016picmemory},  little attention has been paid to creating opportunities for social interactions and promoting social integration in residential care. 
Motivated by recent \textit{needfinding} studies \cite{caforio2017viability,ibarra2017stimulating}, pointing to the potential of reminiscence technology to support NH social activities, we conducted semi-structured interviews and pilot reminiscence sessions to validate the concept of a reminiscence application that builds on the social functions of reminiscence therapy to support (and create opportunities for) co-located and virtual interactions.

\section{Design concept}

The concept, shown in Figure \ref{figure:collage}, resulted from visits, interviews, and focus groups in nursing homes and prior needfinding studies \cite{ibarra2017stimulating}.
A central figure in this process is the helper, who will be in charge of handling the device and assisting older adult residents by guiding the activities described next:

\begin{easylist}
\ListProperties(Hide=100, Hang=true,Style*=\textbullet~)
& \textbf{Digitizing pictures}. The family contributes with pictures related to the resident and with tags (place, date, people), to create an archive that can later be enriched with stories.
Pictures can be added as older adults enter the nursing home, or continuously during their stay.

&  \textbf{Reminiscing and collecting stories}. During visits, family members use pictures to collect stories. The goal is to make visits more interesting, not only to collect information. Therefore, pictures and stories can be revisited at any time, shared with family and friends online, thus generating memories and stimulating conversation. Nursing home staff can also assist or conduct the activity itself.

& \textbf{Engaging in online interactions}. 
Feedback and comments on the stories shared are displayed in a format that facilitates consumption by the resident. Residents can also access pictures and stories from others  (friends) in the NH, potentially leading to face-to-face interactions. 

&  \textbf{Connecting with peers}. The stories and tags are used to find common life points among  residents. These common aspects are used to create mutual awareness, and to provide information about similar interests and affinity with certain topics so as to support animation activities. 
\end{easylist}

The above process describes a social interaction model that aims at stimulating social interactions in nursing homes, via co-located interactions in reminiscence sessions, virtual interactions through shared stories, and facilitating social activities by finding common life points among residents.
The design is built on the opportunities of co-located presence, and in particular, visits by family members and social activities in NHs. The co-located setting poses two main research questions: i) how to effectively support co-located social interactions, and ii) if and how helpers (family members, NH staff, volunteers) can cooperate in the process. 

\section{Methods}

To validate the concept we conducted two separate studies in 4 nursing homes in northern Italy. The studies were approved by the Ethical Committee of the University of Trento.

\begin{marginfigure}[0pc]
  \begin{minipage}{\marginparwidth}
    \centering
  \includegraphics[width=1\columnwidth]{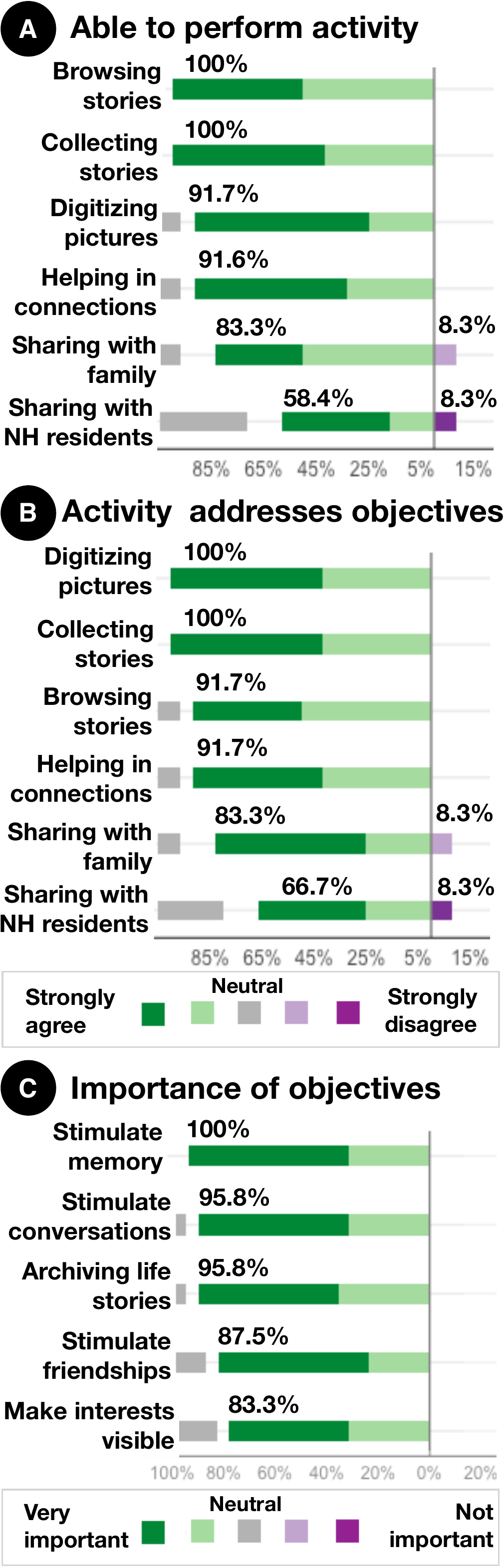}
  \captionof{figure}{Feasibility results}
  \label{figure:feasible}
  \end{minipage}  
\end{marginfigure}

\subsection{Study 1. Evaluating feasibility and perceived value}

In this study we evaluated the willingness and ability of family members (FM) and NH staff to participate in tool-driven reminiscence sessions (see Figure \ref{figure:collage}) as well as the perceived potential of the tool to achieve the intended benefits and the importance of such benefits.
The study was structured as sessions with users in which we presented, via a storyboard, the concept,  main supported activities and expected benefits of the tool. Using a printed mockup (\url{http://invis.io/USB4KFP7N}) we then presented each activity in more detail, requesting feedback on the ability and willingness to perform them and the perceived value (5-point Likert scale).
After evaluating activities, we requested participants to rate (on a 5-point Likert scale) the importance of the objectives addressed by the tool.
For this study we recruited 27 participants (21 females, 15 relatives) but excluded 3 FMs who did not complete the questionnaires.

\subsection{Study 2. Prototyping co-located reminiscence sessions}

We prototyped the experience of co-located social interactions in reminiscence sessions among resident-relative pairs. With the help of the NHs, we recruited 3 diverse resident-relative pairs (see Table~\ref{tab:table1}) who volunteered to participate. 
Prior to the study, FMs were asked to share 10 old pictures from their older adult relatives which we later used to prepare the supporting tool for the reminiscence session.

\begin{marginfigure}[1pc]
  \begin{minipage}{\marginparwidth}
    \centering
  \includegraphics[width=1\columnwidth]{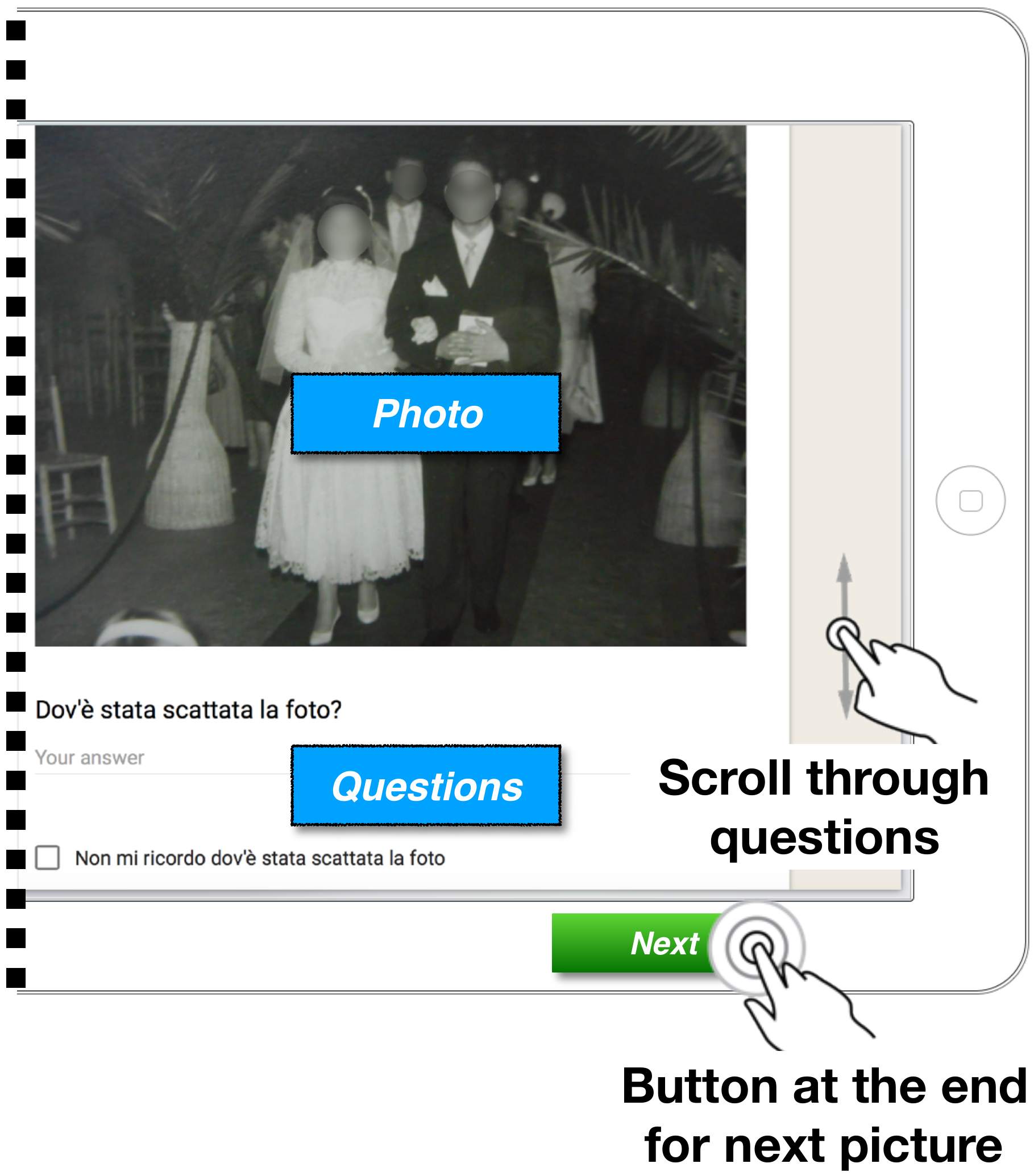}
  \captionof{figure}{Google form running on an iPad tablet, as used in the reminiscence sessions. Annotations highlight main content and gestures. The screen was cut on the left for visibility.}
  \label{figure:form}
  \end{minipage}  
\end{marginfigure}

The session was designed to have FMs as "helpers", using an iPad tablet. 
We used Google Forms to prototype the story collection and reminiscence session (Figure \ref{figure:form}).
The form was designed to emphasize pictures and request picture  information (e.g. place, date, people, activity). 
The session started with demographic information completed by the FM on the tablet, which also served as a training and tutorial on tablet gestures and navigation. The FMs then joined their older relatives in a 25 minutes reminiscence session, where they were asked to go through the pictures and engage their relatives in storytelling while collecting picture information. Two researchers were present, one guided the activities and another observed and took notes.

\section{Results}

\subsection{Feasibility and perceived value}

We evaluated all activities for feasibility and perceived value as agreement with the statements "\textit{I will be able to perform this activity on my own}" and "\textit{This activity helps to accomplish the objectives described}", respectively. 
As seen in Figure \ref{figure:feasible} (A), all participants said they would be able to  "browse" and "collect" stories. 
Only sharing with "family" (83\%) and "NH residents" (58\%) received at least one response score below neutral. 
For perceived value (B), "digitizing" and "collecting" had only positive scores. Again, (virtual) sharing with "family" (83\%) and "other residents" (66\%) had lower (though overall still positive) scores. 

Participants perceived the objectives addressed by the tool as important or very important. As seen in Figure \ref{figure:feasible} (C), this was the case particularly for "stimulating memory" (100\%), conversations and collecting stories.
Making residents aware of the interest generated by their stories (83\%) and fostering bonding among residents (87\%) received a few more neutral scores, all from FMs. NH staff on the other hand saw these objectives as very important. As for the perceived usefulness of the tool for the resident or for themselves, the response was quite high for participants (91\%) but lower for NH residents (71\%). NH staff was generally very positive while some FMs were more skeptical in relation to the usefulness of the tool for their relatives.

\subsection{Co-located reminiscence session}

Overall, participants had a good time during the reminiscence session (GUESS \cite{phan2016development} Enjoyment subscale  6.6/7 for P2 and P3, and 7/7 for P1). Participants did not experience problems in operating the tablet or touch-typing.

\begin{margintable}[1pc]
  \begin{minipage}{\marginparwidth}
    \centering
    \begin{tabular}{r l l l}
     P ID & Sex & Age \\
      \toprule
      FM1 & F & 37 \\
      OA1 & M &  72 (speech)\\
      \midrule
      FM2 & M & 57 &\\
      OA2 & F & 88 (memory)\\
      \midrule
      FM3 & F & 60 &\\
      OA3 & F & 80+\\ 
    \end{tabular}
    \caption{Pilot participants. Impairments indicated in parenthesis. OA3 provided only a range for age.}~\label{tab:table1}
  \end{minipage}
\end{margintable}

Elicitation was mostly prescriptive, heavily based on the form questions, which seem to deter participants from coming up with more engaging questions to generate conversation.
We identified well defined information elicitation strategies: \textit{verification questions} (Who is in the picture? It is <Person>, right? [Yes]), which gave little space for symmetric conversations, \textit{leading questions}, (Who is this? Which aunt?) in which the FM provided clues in a clear effort to facilitate recall, and \textit{recall questions} (Who is this person?), seeking a direct answer from the OA. 
Independently of the strategy, it was clear that the FMs already had the answer to these questions, and in some cases filled them out directly without asking the participant. The use of the tablet in this context switched from \textit{show} and \textit{type}, and behaviors ranged from controlling (FM) to more shared (FM-OA).

The sessions were effective in collecting information. All participants completed information for 4 pictures within the 25 minutes allowed, leaving no fields blank. However, this focus on information collection hindered social interactions. While FMs were typing, OAs were mostly silent. 
This silence occurred both when relatives where discussing the picture first and typing later, and when they where asking for information and typing simultaneously. 

Storytelling emerged naturally, with picture aspects (people, place) triggering specific memories, even when residents were not able to recall the information for that particular picture (e.g., when revealed that <Person> was in the picture, P1 would describe the person "<Person> moved to <City>.. he had 5 children.."). In P2, involving a subject with memory recall issues, the FM would try (unsuccessfully) a question, and then tell the related story to the resident. The resident in this case would not tell stories but leave impressions ("The happiest day of my life").  
Conversations were more natural when pursuing the topic of conversation that triggered a story. Because of the design of the form, participant tended to follow a script which made it less natural. Interestingly, FMs recurred to few \textit{reactions} in response to pictures (e.g., appraisal, curiosity) to engage older adults. 

\section{Discussion and Conclusion}

The results point to a tool that is considered useful and that addresses objectives considered as very important -- in particular stimulating memory and conversations. 
Results also indicate feasibility and willingness of FM and NH staff to play an active role, especially in co-located activities. 

In terms of design, we learned that i) the interface should motivate more symmetric interactions, avoiding scripted conversation, ii) tagging should be separated from the reminiscence session (reduced, or brought up only when necessary), as FMs are able to provide most of the information; iii) the tool should allow for non-linear navigations, following elements of interest based on the flow of conversation.

As ongoing work we are exploring design alternatives to support specific co-located tasks and the use of AI to support story elicitation, to reduce the need for manual tagging, which has recently shown exciting results   \cite{mostafazadeh2016generating,tran2016rich}.

\section{Acknowledgements}
This project has received funding from the EU Horizon 2020 research and innovation programme under the Marie Sk\l{}odowska-Curie grant agreement No 690962. This work was also supported by the ``Collegamenti'' project funded by the Province of Trento (l.p. n.6-December 13th 1999).

\balance{} 

\bibliographystyle{SIGCHI-Reference-Format}
\bibliography{reconect,colocated,ref}

\end{document}